\documentclass[12pt,noshowpacs,nofootinbib,notitlepage,amsmath,amssymb]{revtex4-2}
\usepackage{setspace}
\allowdisplaybreaks
\usepackage{graphicx,color}
\usepackage[colorlinks=true,citecolor=blue,linkcolor=blue,urlcolor=blue]{hyperref}
\usepackage[charter]{mathdesign}
\usepackage{multirow}
\usepackage{enumerate}
\usepackage{array}
\usepackage{booktabs}

\DeclareSymbolFontAlphabet{\mathcal}{symbols}
\DeclareSymbolFont{symbols}{OMS}{xmdcmsy}{m}{n}
\DeclareSymbolFont{largesymbols}{OMX}{cmex}{m}{n}
\SetSymbolFont{symbols}{bold}{OMS}{xmdcmsy}{b}{n}

\begin{document}  
\title{\color{blue}\Large Nonsingular solutions to the Einstein equations\\on piecewise-Lorentzian manifolds}

\author{Bob Holdom}
\email{bob.holdom@utoronto.ca}
\affiliation{Department of Physics, University of Toronto, Toronto, Ontario, Canada  M5S 1A7}
\begin{abstract}
We consider 4-dimensional spacetime manifolds that are piecewise Lorentzian, where the Lorentzian components of the manifold are separated by codimension-one planes (spacelike or timelike) on which the metric is degenerate. Such manifolds are of interest because they enlarge the smooth and nonsingular solution space of the Einstein equations. Planes of degeneracy that are perpendicular to each other can exist simultaneously. We describe various solutions of this type to the vacuum equations $G_{\mu\nu}=0$ and $G_{\mu\nu}+\Lambda g_{\mu\nu}=0$, and to $G_{\mu\nu}= 8\pi G T_{\mu\nu}$ for a perfect fluid. Novel examples include static gravitational lumps of finite curvature and a spacetime that responds to a cosmological constant via oscillations in time and/or space. A spacelike degeneracy plane can be used to avoid the big bang singularity, as we have further described elsewhere.
\end{abstract}

\maketitle 

\section{Introduction}
A basic premise of general relativity is that spacetime is a 4-dimensional Lorentzian manifold, implying that the metric is smooth and nondegenerate everywhere. On the other hand some of the solutions of general relativity that are of the most fundamental interest are plagued by a curvature singularity. We feel that this predicament provides sufficient motivation to step somewhat outside the realm of general relativity. In particular we shall study manifolds that are not Lorentzian everywhere, with the intention of determining whether the enlarged solution space can more readily avoid curvature singularities. We shall show that this is indeed the case, and that there is a large class of such solutions. These solutions to the Einstein equations can be described as being  piecewise Lorentzian. These are new solutions that lie outside the usual scope of solutions in general relativity, as reviewed for example in \cite{stef}.

Our class of metrics involve one or more smooth functions that each depend on a single coordinate. First derivatives of these functions appear in the metric. These metrics have the novel property that the resulting Riemann curvature tensor, and thus also an Einstein tensor, does not involve higher than first derivatives. Another unusual and related property is that these metric functions continue to be arbitrary after solving the Einstein equations. Some of the arbitrariness is related to a form invariance of the metrics under certain coordinate transformations as we shall explain. Some choices of the metric functions may give known solutions of general relativity. But if a metric function has a local extremum then the metric becomes degenerate on a 3-dimensional hypersurface, and we move outside the realm of general relativity. These hypersurface can be spacelike or timelike (with normals that are timelike or spacelike respectively). When spacelike they correspond to all of space at a given instant in time, and when timelike they appear as a static plane of two spatial dimensions. In either case we refer to them as degeneracy planes. There can be an arbitrary number of planes with the same orientation, and planes of orthogonal orientation can be present simultaneously.

An example of one of these new solutions involving spacelike degeneracy planes was recently presented in \cite{Holdom:2023xip}, in the context of cosmology. The universe can move from contraction to expansion, or vice versa, without a singularity, for nonvanishing scale factor and without exotic matter. The whole spacetime is geodesically complete.

When an infinite set of degeneracy planes are equally spaced along some coordinate then this will correspond to a metric that is periodic in that direction. In the context of solutions to $G_{\mu\nu}+\Lambda g_{\mu\nu}=0$, this is of interest because it corresponds to a different way for spacetime to respond to a cosmological constant. These oscillations can have such short time or length scales so as to be effectively unobservable, leaving a spacetime that appears to be flat. We shall see that these oscillations can be occurring with respect to 1, 2, 3 or 4 spacetime directions simultaneously.

Another novel type of solution is one that looks like a gravitational lump. This is a nonsingular static solution that is asymptotically flat and that has spherically symmetric curvature invariants. The metric itself is not spherically symmetric and it involves three perpendicular and intersecting degeneracy planes at fixed $x$, $y$ and $z$ respectively. These solutions exist as vacuum solutions to $G_{\mu\nu}=0$ or as perfect fluid solutions to $G_{\mu\nu}= 8\pi G T_{\mu\nu}$. In the latter case the standard energy conditions are satisfied but the pressure $p=w\rho$ must be mildly negative with $-\frac{1}{7}<w<0$.

With the motivation to increase the smooth and nonsingular solution space of the Einstein equations via piecewise-Lorentzian manifolds, and with the knowledge that there are physically interesting examples of such solutions, we thus proceed to study some of the details of this class of metrics. We describe the metrics and their symmetries in the next two sections. In the subsequent three sections we describe the various solutions. In the final section we step back and comment on some of the physics of the degeneracy planes, such as the behavior of geodesics.

\section{The class of metrics}\label{s1}
Our class of metrics is as listed below. We have Cartesian-like coordinates $t$, $x$, $y$, $z$ and dimensionless coordinates are defined as $\bar t=t/\ell$, $\bar x=x/\ell$, $\bar y=y/\ell$, $\bar z=z/\ell$ for some length scale $\ell$. The metrics involve one or more of the functions $a(\bar x),b(\bar y),c(\bar z),d(\bar t)$ and one or more of the constants $A,B,C,D$. We require that these functions and constants be positive. They are also all dimensionless. There are also one or two constant exponents $u$ and $v$. Upon inspection of these metrics the reader will notice the patterns that distinguish the different types.

Type 1a
\begin{align}
\textit{ds}^2&=-\Sigma \textit{dt}^2+A a'(\bar x)^2\Sigma^u \textit{dx}^2+\Sigma \textit{dy}^2+\Sigma \textit{dz}^2\nonumber\\
\Sigma&= a(\bar x)\nonumber
\end{align}

Type 1b
\begin{align}
\textit{ds}^2&=-D d'(\bar t)^2\Sigma^u \textit{dt}^2+\Sigma \textit{dx}^2+\Sigma \textit{dy}^2+\Sigma \textit{dz}^2\nonumber\\
\Sigma&= d(\bar t)\nonumber
\end{align}

Type 2a
\begin{align}
\textit{ds}^2&=-\Sigma^v \textit{dt}^2+A a'(\bar x)^2\Sigma^u \textit{dx}^2+B b'(\bar y)^2 \Sigma^u \textit{dy}^2+\Sigma^v \textit{dz}^2\nonumber\\
\Sigma&= a(\bar x)+b(\bar y)\nonumber
\end{align}

Type 2b
\begin{align}
\textit{ds}^2&=-D d'(\bar t)^2 \Sigma^u \textit{dt}^2+A a'(\bar x)^2\Sigma^u \textit{dx}^2+ \Sigma^v \textit{dy}^2+\Sigma^v \textit{dz}^2\nonumber\\
\Sigma&= a(\bar x)+d(\bar t)\nonumber
\end{align}

Type 3a
\begin{align}
\textit{ds}^2&=-\Sigma^v \textit{dt}^2+A a'(\bar x)^2\Sigma^u \textit{dx}^2+B b'(\bar y)^2 \Sigma^u \textit{dy}^2+C c'(\bar z)^2\Sigma^u \textit{dz}^2\nonumber\\
\Sigma&= a(\bar x)+b(\bar y)+c(\bar z)\nonumber
\end{align}

Type 3b
\begin{align}
\textit{ds}^2&=-D d'(\bar t)^2\Sigma^u \textit{dt}^2+A a'(\bar x)^2\Sigma^u \textit{dx}^2+B b'(\bar y)^2 \Sigma^u \textit{dy}^2+\Sigma^v \textit{dz}^2\nonumber\\
\Sigma&= a(\bar x)+b(\bar y)+d(\bar t)\nonumber
\end{align}

Type 4
\begin{align}
\textit{ds}^2&=-D d'(\bar t)^2\Sigma^u \textit{dt}^2+A a'(\bar x)^2\Sigma^u \textit{dx}^2+B b'(\bar y)^2 \Sigma^u \textit{dy}^2+C c'(\bar z)^2\Sigma^u \textit{dz}^2\nonumber\\
\Sigma&= a(\bar x)+b(\bar y)+c(\bar z)+d(\bar t)\nonumber
\end{align}

The presence of the derivatives\footnote{Primes denote derivatives with respect to the barred argument of the function.} indicate that the functions $a,b,c,d$ cannot be constants. With derivatives in the metrics, the Christoffel symbols can and do depend on second derivatives. Thus the calculation of the Riemann tensor can involve third derivatives, but such terms and terms with second derivatives all cancel, leaving the components of the Riemann tensor to depend on just first derivatives.

For Type 1 the Einstein tensor is diagonal, but it is not automatically diagonal for the other types. The off-diagonal elements are proportional to $2u-v+2$ for Type 2 and to $u^2 + 2 v u - v^2 + 2 u + 2 v$ for Type 3 and to $2+u$ for Type 4. Requiring that these expressions vanish leaves one degree of freedom for Type 2 and Type 3, as is the case for Type 1, while Type 4 has no remaining freedom. Once the Einstein tensor is diagonal then both it and the metric tensor depend on the same squares of first derivatives in the same components. Using the freedom in the choice of the exponents $u$ and $v$ can then yield solutions.

These are solutions for any (nonconstant) choice of the functions $a,b,c,d$. But the meaning of this should be considered in the context of coordinate transformations. Our class of metrics maintain their form under coordinate transformations such as $\bar x=f(\bar x_{\rm new})$, for some one-to-one function $f$, for each coordinate on which the metric depends. That is, the effect of such a coordinate transformation is to simply replace $a(\bar x)$ by $a_{\rm new}(\bar x_{\rm new})= a(f(\bar x_{\rm new}))$ in the transformed metric, with analogous replacements possible for $b$, $c$ and $d$. This form invariance is related to the presence of the squares of derivatives in the metric, where the derivatives in the new metric are with respect to the new barred coordinates.\footnote{The transformation of the derivatives absorbs the multiplicative factors in the coordinate transformation of the metric tensor.} This freedom to trade $a(\bar x)$ for $a_{\rm new}(\bar x_{\rm new})$ (and similarly for $b(\bar y)$, $c(\bar z)$, $d(\bar t)$) via a coordinate transformation makes it less peculiar for there to be solutions for any choice of $a,b,c,d$. 

But not all aspects of the functions $a,b,c,d$ can be altered by these coordinate transformations. Of interest to us are any local extrema of these functions, since these mark the degeneracy planes. Both the number of these extrema and the values of the metric functions at these extrema are left invariant by the coordinate transformations. These quantities are free to choose, and each such choice corresponds to a physically different spacetime.

\section{Symmetries}\label{s2}
Symmetries of a metric are described in terms of isometries and homotheties \cite{stef}, and Table \ref{t1} lists the associated Killing and homothetic vectors for our class of metrics for $u\neq-2$ and arbitrary $v$. Whereas a Killing vector $\xi$ is defined by a vanishing Lie derivative of the metric ${\cal L}_\xi g_{\mu\nu}=0$, any homothetic vector $\zeta$ in the table satisfies
\begin{align}
{\cal L}_\zeta g_{\mu\nu}=2(2+u)g_{\mu\nu},\quad {\cal L}_\zeta g^{\mu\nu}=-2(2+u)g^{\mu\nu}
.\label{e17}\end{align}
It also turns out that
\begin{align}
{\cal L}_\zeta \Sigma=2\Sigma
,\label{e16}\end{align}
where $\Sigma$ is defined according to the metric type. Homothetic vectors in general satisfy
\begin{align}
{\cal L}_\zeta R^\mu_{\hspace{1ex}\alpha\beta\gamma}=0,\quad {\cal L}_\zeta (\nabla_\rho R^\mu_{\hspace{1ex}\alpha\beta\gamma})=0,
\label{e18}\end{align}
where the latter is also true for more covariant derivatives. These results depend on the canonical placement of indices as shown. Since $R_{\mu\nu}$ is obtained by contracting an upper and lower index on $R^\mu_{\hspace{1ex}\alpha\beta\gamma}$, we also have ${\cal L}_\zeta R_{\mu\nu}=0$, and then also for the Einstein tensor ${\cal L}_\zeta G_{\mu\nu}=0$.

\renewcommand{\arraystretch}{2}
\setlength{\tabcolsep}{0.5em}
\begin{table}[htbp]
\centering
\begin{tabular}{ |c|c|c| } 
\hline
\multirow{2}{*}{Type 1a} & Killing & $\partial_t,\;\;\partial_y,\;\;\partial_z,\;\;y\partial_t+t\partial_y,\;\;z\partial_t+t\partial_z,\;\;y\partial_z-z\partial_y\;$ \\
\cline{2-3}
& homothetic & $(1+u)t\partial_t+2\frac{a(x)}{a'(x)}\partial_x+(1+u)y\partial_y+(1+u)z\partial_z$ \\  
\hline
\multirow{2}{*}{Type 1b} & Killing & $\partial_x,\;\;\partial_y,\;\;\partial_z,\;\;y\partial_x-x\partial_y,\;\;z\partial_x-x\partial_z,\;\;y\partial_z-z\partial_y\;$  \\
\cline{2-3}
& homothetic & $2\frac{d(t)}{d'(t)}\partial_t+(1+u)x\partial_x+(1+u)y\partial_y+(1+u)z\partial_z$ \\  
\hline
\multirow{2}{*}{Type 2a} & Killing & $\partial_t,\;\;\partial_z,\;\;z\partial_t+t\partial_z,\;\;\frac{1}{a'(x)}\partial_x-\frac{1}{b'(y)}\partial_y$\\ 
\cline{2-3}
& homothetic & $(2+u-v)t\partial_t+2\frac{a(x)}{a'(x)}\partial_x+2\frac{b(y)}{b'(y)}\partial_y+(2+u-v)z\partial_z$ \\  
\hline
\multirow{2}{*}{Type 2b} & Killing & $\partial_y,\;\;\partial_z,\;\;z\partial_y-y\partial_z,\;\;\frac{1}{a'(x)}\partial_x-\frac{1}{d'(t)}\partial_t$\\ 
\cline{2-3}
& homothetic & $2\frac{d(t)}{d'(t)}\partial_t+2\frac{a(x)}{a'(x)}\partial_x+(2+u-v)y\partial_y+(2+u-v)z\partial_z$ \\
\hline
\multirow{3}{*}{Type 3a} & \multirow{2}{*}{Killing} & $\partial_t,\;\;\frac{1}{a'(x)}\partial_x -\frac{1}{b'(y)}\partial_y,\;\;\frac{1}{a'(x)}\partial_x -\frac{1}{c'(z)}\partial_z,$\\
&&$\frac{B b(y)-C c(z)}{a'(x)}\partial_x+\frac{C c(z)-A a(x)}{b'(y)}\partial_y+\frac{A a(x)-B b(y)}{c'(z)}\partial_z$\\ 
\cline{2-3}
& homothetic & $(2+u-v)t\partial_t+2\frac{a(x)}{a'(x)}\partial_x+2\frac{b(y)}{b'(y)}\partial_y+2\frac{c(z)}{c'(z)}\partial_z$ \\  
\hline
\multirow{3}{*}{Type 3b} & \multirow{2}{*}{Killing} & $\partial_z,\;\;\frac{1}{b'(y)}\partial_y -\frac{1}{d'(t)}\partial_t,\;\;\frac{1}{b'(y)}\partial_y -\frac{1}{c'(z)}\partial_z,$ \\
&&$\frac{B b(y)-A a(x)}{d'(t)}\partial_t-\frac{D d(t)+B b(y)}{a'(x)}\partial_x+\frac{A a(x)+D d(t)}{cb'(y)}\partial_y$\\ 
\cline{2-3}
& homothetic & $2\frac{d(t)}{d'(t)}\partial_t+2\frac{a(x)}{a'(x)}\partial_x+2\frac{b(y)}{b'(y)}\partial_y+(2+u-v)z\partial_z$ \\
\hline
\multirow{2}{*}{Type 4} & Killing & six not shown \\
\cline{2-3}
& homothetic & $2\frac{d(t)}{d'(t)}\partial_t+2\frac{a(x)}{a'(x)}\partial_x+2\frac{b(y)}{b'(y)}\partial_y+2\frac{c(z)}{c'(z)}\partial_z$ \\
\hline
\end{tabular}
\caption{A listing of the Killing and homothetic vectors for each metric type when $u\neq-2$. The bars on the dimensionless coordinates $\bar x,\bar y,\bar z,\bar t$ are suppressed in this table.}
\label{t1}
\end{table}

Curvature invariants can be constructed by taking some numbers of Riemann tensors and covariant derivatives, with all indices different and in canonical placement, along with some number $N_i$ of inverse metrics, such that all indices can be contracted. For example consider curvature invariants constructed from one, two or three Riemann tensors or, for example, from three Riemann tensors and two covariant derivatives. We have $N_i=1,2,3,4$ respectively. From (\ref{e17}) and (\ref{e18}), the number $N_i$ determines the transformation property of the curvature invariant,
\begin{align}
{\cal L}_\zeta \mbox{[curvature invariant]}=-2N_i(2+u)\mbox{[curvature invariant]}
.\end{align}
We note that $\Sigma^{-2-u}$ has the same transformation property as the inverse metric, given (\ref{e16}). Thus we may expect that any curvature invariant will have a simple dependence on $\Sigma$ as follows
\begin{align}
\mbox{[curvature invariant]}\propto \frac{1}{\Sigma^{N_i(2+u)}}
.\label{e11}\end{align}
This turns out to be the case by direct calculation. Curvature invariants depend on the functions $a,b,c,d$ only in this way and derivatives of these functions do not appear.

Curvature invariants are constant in the special case of $u=-2$, according to (\ref{e11}). In fact we have solutions of the equation $G_{\mu\nu}+\Lambda g_{\mu\nu}=0$ in this case (with $v=-2$ required as well for Type 2 and 3 metrics). This equation does not allow $g_{\mu\nu}$ to transform in a (nontrivial) homothetic way since both $G_{\mu\nu}$ and $\Lambda$ do not, and this ties in with the absence of a homothetic vector when $u=-2$ according to (\ref{e17}). For these solutions to $G_{\mu\nu}+\Lambda g_{\mu\nu}=0$ the number of Killing vectors is 10.

Finally we note that the discussion in this section only applies away from the degeneracy planes; on these planes a derivative vanishes and some of the Killing or homothetic vectors become ill-defined. A component of the metric vanishes and the remaining symmetries are those of the lower-dimensional metric.

\section{Type 1}\label{s5}
We start by recovering some standard solutions of general relativity. For the Type 1a metric, with $u=-2$, the choice 
\begin{align}
a(\bar x)=\frac{\alpha^2}{x^2},\quad \alpha=2\sqrt{A} \ell,
\label{e6}\end{align}
yields
\begin{align}
\textit{ds}^2&=\frac{\alpha^2}{x^2}(-\textit{dt}^2+\textit{dx}^2+\textit{dy}^2+\textit{dz}^2)
.\end{align}
With $x>0$ this is the standard half covering of the anti-de Sitter (AdS) surface in 5D, as defined by $X_1^2+X_2^2+X_3^2-T_1^2-T_2^2=-\alpha^2$. $\alpha$ is the radius of curvature.

For the Type 1b metric, also for $u=-2$, the choice
\begin{align}
d(\bar t)=\frac{\alpha^2}{t^2},\quad \alpha=2\sqrt{D} \ell
,\label{e7}\end{align}
yields
\begin{align}
\textit{ds}^2&=\frac{\alpha^2}{t^2}(-\textit{dt}^2+\textit{dx}^2+\textit{dy}^2+\textit{dz}^2)
.\label{e24}\end{align}
With $t>0$ this is the standard half covering of the de Sitter (dS) surface in 5D, as defined by $X_1^2+X_2^2+X_3^2+X_4^2-T_2^2=\alpha^2$. If we instead take $d(\bar t)=e^{2t/\alpha}$ then we get
\begin{align}
\textit{ds}^2&=-\textit{dt}^2+e^{2t/\alpha}(\textit{dx}^2+\textit{dy}^2+\textit{dz}^2)
,\end{align}
another flat slicing of the dS surface.

These examples involve functions $a$ and $d$ that have certain properties, such as their domain and ranges and that they are one-to-one. But these particular properties are not necessary for the Type 1 metrics to provide solutions of Einstein equations. All that is needed is the choice of $u$. To see this more clearly we give the nonzero elements of the Einstein tensor, 
\begin{align}
\textrm{Type 1a}:\quad &G_\textit{xx}=\frac{3 }{4 \ell^2}\frac{a'(\bar x)^2}{a(\bar x)^2}\quad G_\textit{tt}=-G_\textit{yy}=-G_\textit{zz}=\frac{(1 + 2u)}{4A\ell^2}\frac{1}{a(\bar x)^{1 + u}},\nonumber\\
\textrm{Type 1b}:\quad &G_\textit{tt}=\frac{3 }{4 \ell^2}\frac{d'(\bar t)^2}{d(\bar t)^2}\quad G_\textit{xx}=G_\textit{yy}=G_\textit{zz}=\frac{(1 + 2u)}{4D\ell^2}\frac{1}{d(\bar t)^{1 + u}}.\label{e10}
\end{align}
There are no second derivatives and the squares of first derivatives occurs in the same components as they do in the metric. For the choice $u=-2$ it is then simple to see a solution to $G_{\mu\nu}+\Lambda g_{\mu\nu}=0$ for the Type 1a metric when $\Lambda=-3/\alpha^2$ for $\alpha$ in (\ref{e6}), and for the Type 1b metric when $\Lambda=3/\alpha^2$ for $\alpha$ in (\ref{e7}). Other than being positive, the functions $a(\bar x)$ or $d(\bar t)$ are not constrained.

We can thus consider functions $a(\bar x)$ or $d(\bar t)$ having local extrema and thereby describe a piecewise-Lorentzian manifold. The curvature invariants are still the same as for the standard AdS or dS solutions above, but the global properties certainly differ. An interesting case is when $a(\bar x)$ or $d(\bar t)$ are oscillating and periodic. We shall say more about this case in the final section. Our results for the Type 2, 3 and 4 metrics in the next section indicate that this type of picture can be extended to multiple dimensions simultaneously.

Away from $u=-2$, the Type 1b metric provides solutions to $G_{\mu\nu}= 8\pi G T_{\mu\nu}$ for a perfect fluid. These solutions are the topic of \cite{Holdom:2023xip} where cosmologies of the bounce and oscillating types are presented, as well as various generalizations.

\section{Types 2, 3, 4}\label{s3}
\subsection{Vacuum solutions to $G_{\mu\nu}+\Lambda g_{\mu\nu}=0$}\label{s3a}

The vacuum equations $G_{\mu\nu}+\Lambda g_{\mu\nu}=0$ are solved by the Type 2 and Type 3 metrics when $u=v=-2$ and by the Type 4 metric when $u=-2$. These solutions have the expected geometric properties in terms of the curvature invariants
\begin{align}
R_{\mu\nu}&=\Lambda g_{\mu\nu}=-{\rm sign}(q)\frac{3}{\alpha^2}g_{\mu\nu},\quad \alpha^2=\ell^2/|q|,\nonumber\\
R&=-{\rm sign}(q)\frac{12}{\alpha^2},\quad R_{\mu\nu\rho\sigma}R^{\mu\nu\rho\sigma}=\frac{24}{\alpha^4},
\label{e3}\end{align}
and by having 10 Killing vectors. The Weyl tensor vanishes in all cases. $q$ is defined for the different types as
\begin{align}
\mbox{Type 2a}\quad q&=\frac{A+B}{AB},\label{e1}\\
\mbox{Type 2b}\quad q&=\frac{D-A}{DA},\label{e2}\\
\mbox{Type 3a}\quad q&=\frac{AB+AC+BC}{ABC},\label{e4}\\
\mbox{Type 3b}\quad q&=\frac{AD+BD-AB}{ABD},\label{e5}\\
\mbox{Type 4}\quad q&=\frac{ABD+ACD+BCD-ABC}{ABCD}.
\end{align}
When $q>0$ (or $q<0$) we have the AdS (or dS) geometry with radius of curvature $\alpha$. Since the constants $A,B,C,D$ are all positive, we see that $q$ for Types 2b, 3b and 4 can be positive or negative while $q$ is strictly positive for Types 2a and 3a. The functions $a,b,c,d$ are still free to choose.  Once again they can be oscillatory, corresponding to oscillations along 2, 3 or 4 space and/or time directions simultaneously. $\ell$ sets the length scale or the timescale of the oscillation, and $\ell$ in turn is related to the cosmological constant.

\subsection{Vacuum solutions to $G_{\mu\nu}=0$}\label{s3b}

The vacuum equations $G_{\mu\nu}=0$ are solved by the Type 2 metrics when $u=-1/2$ and $v=1$. The curvature invariants are
\begin{align}
R_{\mu\nu\rho\sigma}R^{\mu\nu\rho\sigma}&=C_{\mu\nu\rho\sigma}C^{\mu\nu\rho\sigma}=\frac{3q^2}{4\ell^4\Sigma^{3}}
.\label{e14}\end{align}
For Type 2a, $\Sigma= a(\bar x)+b(\bar y)$ and the value of $q$ is given in (\ref{e1}), while for Type 2b, $\Sigma= a(\bar x)+d(\bar t)$ and the value of $q$ is given in (\ref{e2}).

The vacuum equations $G_{\mu\nu}=0$ are solved by the Type 3 metrics when $u=4$ and $v=-2$. Here the curvature invariants are
\begin{align}
R_{\mu\nu\rho\sigma}R^{\mu\nu\rho\sigma}&=C_{\mu\nu\rho\sigma}C^{\mu\nu\rho\sigma}=\frac{192q^2}{\ell^4\Sigma^{12}}
.\label{e15}\end{align}
For Type 3a, $\Sigma= a(\bar x)+b(\bar y)+c(\bar z)$ and the value of $q$ is given in (\ref{e4}), while for Type 3b, $\Sigma= a(\bar x)+b(\bar y)+d(\bar t)$ and the value of $q$ is given in (\ref{e5}).

These Type 2 and 3 vacuum solutions are of Petrov type D. The relation of the power of $\Sigma$ in (\ref{e14}) and (\ref{e15}) to the value of $u$ agrees with the homothety arguments in Sec.~\ref{s2}. The curvature invariants avoid singularities as long as the range of $\Sigma$ does not include zero. When the functions $a,b,c,d$ are oscillatory, the curvature invariants are oscillatory along two dimensions (Type 2) or three dimensions (Type 3). For the Types 2b and 3b, one of these dimensions is time.

For Type 3a it is interesting to consider functions $a,b,c$ that are not oscillating, and in particular when they are such that $\Sigma=s+\bar x^2+\bar y^2+\bar z^2$ for a constant $s>0$. Then the invariants in (\ref{e15}) are $\propto (s+\bar r^2)^{-12}$ in spherical coordinates. The solution rapidly approaches asymptotic flatness and there is no curvature singularity at the origin. This is the spherical gravitational lump mentioned in the Introduction. The metric itself lacks spherical symmetry, as illustrated by the volume element,
\begin{align}
\sqrt{-g}=\frac{8}{\ell^{13}}\sqrt{ABC}|x y z|(s+\bar r^2)^5.
\label{e19}\end{align}

The analogous example for Type 2a is $\Sigma=s+\bar x^2+\bar y^2$ which gives a lump of cylindrical geometry of infinite extent in the $z$ direction. The corresponding Type 3b and Type 2b solutions are time dependent lumps with curvature invariants smoothly turning on and then off as a function of time; one is a spatially infinite tube while the other is a spatially infinite wall. One can also make the choice $\Sigma=s+\bar x^2+\bar y^2+\bar z^2$ for the Type 3a solution of $G_{\mu\nu}+\Lambda g_{\mu\nu}=0$, but in this case the invariants are of course constant.

\subsection{Perfect fluid solution}\label{s3c}
The Type 3a metric provides a perfect fluid solution to $G_{\mu\nu}= 8\pi G T_{\mu\nu}$. In this case the energy density depends on the spatial coordinates,
\begin{align}
\rho(\bar x,\bar y,\bar z)=\frac{1}{8\pi G\ell^2}q\frac{u(4-u)}{4}\frac{1}{\Sigma^{2+u}},\quad \Sigma=a(\bar x)+b(\bar y)+c(\bar z)
.\end{align}
The value of $q$ is given in (\ref{e4}). We see that $\rho>0$ for $0<u<4$. But there are two possible values of the equation state parameter in $p=w\rho$ because there are two possible values of $v$ for a given $u$ that produce a diagonal Einstein tensor,
\begin{align}
v=u + 1 \pm \sqrt{2u^2 + 4u + 1}
.\end{align}
Correspondingly
\begin{align}
w=\frac{2+3u\pm2\sqrt{2 u^{2}+4 u +1}}{4-u}
.\end{align}

The plus branch has $w$ varying monotonically in $1<w<\infty$ as $u$ varies in $0<u<4$. This violates the dominant energy condition. $u=4$ gives $v=12$, $\rho=0$ and $p>0$, while $u=0$ gives $v=2$, $\rho=p=0$ and a flat spacetime. This branch also has a $w=1/3$ solution ($u=-\frac{2}{7},v=\frac{6}{7}$) but then $\rho<0$.

The minus branch is more interesting. Now the standard energy conditions are satisfied as $w$ varies monotonically in $0>w>-\frac{1}{7}$ as $u$ varies in $0<u<4$. $u=4$ gives back the Type 3a vacuum solution in Sec.~\ref{s3b} while $u=0$ gives $v=0$, $\rho=p=0$ and a flat spacetime. The minus branch at $u=-2$ produces the Type 3a AdS solution in Sec.~\ref{s3a}.

The curvature invariants including $R$ are in general nonvanishing, but we just give the following result, which is particularly simple and is true for both branches,
\begin{align}
C_{\mu\nu\rho\sigma}C^{\mu\nu\rho\sigma}=\frac{q^2u^2(2+u)^2}{3\ell^4\Sigma^{4+2u}}
.\end{align}
For these perfect fluid solutions, both oscillating solutions and asymptotically-flat solutions can again be considered.

\section{Variations}
One of the defining characteristics of our class of diagonal metrics is that some of the components involve a square of a derivative. We can keep this basic structure intact while allowing more choice for the power of $\Sigma$ that appears in each component. This will typically reduce the symmetries, but additional solutions emerge. The Riemann tensor still has no more than first derivatives. This type of variation for the Type 1 metrics leads us into an extension of the Kasner metrics as discussed in \cite{Holdom:2023xip}. In the remainder of this section we consider this type of variation for the Type 2, 3 and 4 metrics.

\subsection{More vacuum solutions to $G_{\mu\nu}=0$}
A variation of the Type 2 metrics leads to solutions to the vacuum equations $G_{\mu\nu}=0$ that still have a degree of freedom in the choice of the exponents. That is, the following metrics are solutions for any choice of $u$,
\begin{align}
\textit{ds}^2&=-\Sigma^u \textit{dt}^2+A a'(\bar x)^2\Sigma^{u(u-2)/2} \textit{dx}^2+B b'(\bar y)^2 \Sigma^{u(u-2)/2} \textit{dy}^2+\Sigma^{2-u} \textit{dz}^2,\nonumber\\
\Sigma&= a(\bar x)+b(\bar y),
\end{align}
\begin{align}
\textit{ds}^2&=-D d'(\bar t)^2 \Sigma^{u(u-2)/2} \textit{dt}^2+A a'(\bar x)^2\Sigma^{u(u-2)/2} \textit{dx}^2+ \Sigma^u \textit{dy}^2+\Sigma^{2-u} \textit{dz}^2,\nonumber\\
\Sigma&= a(\bar x)+d(\bar t)
.\end{align}
The values of the curvature invariant $R_{\mu\nu\rho\sigma}R^{\mu\nu\rho\sigma}=C_{\mu\nu\rho\sigma}C^{\mu\nu\rho\sigma}$ for these two metrics are, respectively,
\begin{align}
&\frac{ \left(A +B \right)^{2} u^{2} \left(2-u\right)^{2} \left(4-2 u +u^{2}\right)}{4 B^{2} A^{2} \ell^{4}\Sigma^{4-2u+u^{2}}},\quad \Sigma= a(\bar x)+b(\bar y),\\
&\frac{ \left(A -D \right)^{2} u^{2} \left(2-u\right)^{2} \left(4-2 u +u^{2}\right)}{4 D^{2} A^{2} \ell^{4}\Sigma^{4-2u+u^{2}}},\quad \Sigma= a(\bar x)+d(\bar t).
\end{align}
Since $4-2u+u^{2}>0$, these vacuum solutions are nonsingular as long as there are no points where $\Sigma=0$. Results are symmetric under interchange of $u$ with $2-u$. $u=1$ gives back the Type 2a and 2b vacuum solutions in Sec.~\ref{s3b} while $u=2,0$ gives flat space. $u=4,-2$ gives a solution with four Killing vectors, as with $u=1$, while solutions for other values of $u$ have three Killing vectors. A homothety vector exists in all cases and can be used to obtain the power of $\Sigma$ in the curvature invariants as before.

\subsection{More perfect fluid solutions}
A variation of the Type 2a and 3a metrics leads to additional perfect fluid solutions to $G_{\mu\nu}= 8\pi G T_{\mu\nu}$, as follows:
\begin{align}
\textit{ds}^2&=-\Sigma^{2-u} \textit{dt}^2+A a'(\bar x)^2\Sigma^{u(u-2)/2} \textit{dx}^2+B b'(\bar y)^2 \Sigma^u \textit{dy}^2+\Sigma^{u(u-2)/2} \textit{dz}^2,\nonumber\\
\Sigma&= a(\bar x)+b(\bar y),
\end{align}
\begin{align}
\textit{ds}^2&=-\Sigma^{2-u} \textit{dt}^2+A a'(\bar x)^2\Sigma^{u(u-2)/2} \textit{dx}^2+B b'(\bar y)^2 \Sigma^u \textit{dy}^2+C c'(\bar z)^2\Sigma^{u(u-2)/2} \textit{dz}^2,\nonumber\\
\Sigma&= a(\bar x)+b(\bar y)+c(\bar z)
.\end{align}
The energy density is
\begin{align}
\rho=\frac{1}{8\pi G\ell^2}\frac{u (3 u^{3}-16 u^{2}+12 u +16)}{16B}\frac{1}{\Sigma^{2+u}}
,\end{align}
where $\Sigma=a(\bar x)+b(\bar y)$ or $\Sigma=a(\bar x)+b(\bar y)+c(\bar z)$ respectively. For both cases
\begin{align}
p=w\rho,\quad w=\frac{2-u}{2+3u}
.\end{align}
$\rho$ is positive in the ranges $2<u<4$ and $-\frac{2}{3}<u<0$. For the first range $w$ varies in $0>w>-\frac{1}{7}$, while for the second $\infty>w>1$. These ranges match those found for the perfect fluid solution in Sec.~\ref{s3c}. The curvature invariants are more complicated and have various terms with different dependence on $\Sigma$, as may be expected since no homothety vector exists.

\subsection{Nonconformally-flat Einstein spaces}
There is a variation of the Type 2, 3 and 4 metrics that produces a different type of solution, namely solutions to $G_{\mu\nu}+\Lambda g_{\mu\nu}=0$ that are not conformally flat (nonvanishing Weyl tensor). These solutions give rise to
\begin{align}
R_{\mu\nu}=-\frac{12q_1}{\ell^2}g_{\mu\nu},\quad  R_{\mu\nu\rho\sigma}R^{\mu\nu\rho\sigma}&=\frac{384q_2^2}{\ell^4}+C_{\mu\nu\rho\sigma}C^{\mu\nu\rho\sigma},\quad C_{\mu\nu\rho\sigma}C^{\mu\nu\rho\sigma}=\frac{192q_3^2}{\ell^4\Sigma^{12}}.
\label{e8}\end{align}
The invariants are not constant. The following metrics produce such solutions, where we also give the values of $q_1$, $q_2$ and $q_3$ appearing in (\ref{e8}).
\begin{align}
\textit{ds}^2&=-\Sigma^4 \textit{dt}^2+A a'(\bar x)^2\Sigma^{-2} \textit{dx}^2+B b'(\bar y)^2 \Sigma^4 \textit{dy}^2+\Sigma^4 \textit{dz}^2\\
\Sigma&= a(\bar x)+b(\bar y),\quad q_1=\frac{1}{A},\quad q_2=\frac{1}{B},\quad q_3=\frac{1}{A}\nonumber
\end{align}
\begin{align}
\textit{ds}^2&=-D d'(\bar t)^2 \Sigma^{-2} \textit{dt}^2+A a'(\bar x)^2\Sigma^4 \textit{dx}^2+ \Sigma^4 \textit{dy}^2+\Sigma^4 \textit{dz}^2\\
\Sigma&= a(\bar x)+d(\bar t),\quad q_1=-\frac{1}{D},\quad q_2=\frac{1}{D},\quad q_3=\frac{1}{A}\nonumber
\end{align}
\begin{align}
\textit{ds}^2&=-D d'(\bar t)^2 \Sigma^4 \textit{dt}^2+A a'(\bar x)^2\Sigma^{-2} \textit{dx}^2+ \Sigma^4 \textit{dy}^2+\Sigma^4 \textit{dz}^2\\
\Sigma&= a(\bar x)+d(\bar t),\quad q_1=\frac{1}{A},\quad q_2=\frac{1}{A},\quad q_3=\frac{1}{D}\nonumber
\end{align}
\begin{align}
\textit{ds}^2&=-\Sigma^4 \textit{dt}^2+A a'(\bar x)^2\Sigma^{-2} \textit{dx}^2+B b'(\bar y)^2 \Sigma^4 \textit{dy}^2+C c'(\bar z)^2\Sigma^4 \textit{dz}^2\\
\Sigma&= a(\bar x)+b(\bar y)+c(\bar z),\quad q_1=\frac{1}{A},\quad q_2=\frac{1}{A},\quad q_3=\frac{B+C}{BC}\nonumber
\end{align}
\begin{align}
\textit{ds}^2&=-D d'(\bar t)^2\Sigma^{-2} \textit{dt}^2+A a'(\bar x)^2\Sigma^4 \textit{dx}^2+B b'(\bar y)^2 \Sigma^4 \textit{dy}^2+\Sigma^4 \textit{dz}^2\\
\Sigma&= a(\bar x)+b(\bar y)+d(\bar t),\quad q_1=\frac{1}{A},\quad q_2=\frac{1}{A},\quad q_3=\frac{D-B}{BD}\nonumber
\end{align}
\begin{align}
\textit{ds}^2&=-D d'(\bar t)^2\Sigma^4 \textit{dt}^2+A a'(\bar x)^2\Sigma^{-2} \textit{dx}^2+B b'(\bar y)^2 \Sigma^4 \textit{dy}^2+\Sigma^4 \textit{dz}^2\\
\Sigma&= a(\bar x)+b(\bar y)+d(\bar t),\quad q_1=-\frac{1}{D},\quad q_2=\frac{1}{D},\quad q_3=\frac{A+B}{AB}\nonumber
\end{align}
\begin{align}
\textit{ds}^2&=-D d'(\bar t)^2\Sigma^{-2} \textit{dt}^2+A a'(\bar x)^2\Sigma^4 \textit{dx}^2+B b'(\bar y)^2 \Sigma^4 \textit{dy}^2+C c'(\bar z)^2\Sigma^4 \textit{dz}^2\\
\Sigma&= a(\bar x)+b(\bar y)+c(\bar z)+d(\bar t),\quad q_1=-\frac{1}{D},\quad q_2=\frac{1}{D},\quad q_3=\frac{AB+AC+BC}{ABC}\nonumber
\end{align}
\begin{align}
\textit{ds}^2&=-D d'(\bar t)^2\Sigma^4 \textit{dt}^2+A a'(\bar x)^2\Sigma^{-2} \textit{dx}^2+B b'(\bar y)^2 \Sigma^4 \textit{dy}^2+C c'(\bar z)^2\Sigma^4 \textit{dz}^2\\
\Sigma&= a(\bar x)+b(\bar y)+c(\bar z)+d(\bar t),\quad q_1=\frac{1}{A},\quad q_2=\frac{1}{A},\quad q_3=\frac{BD+CD-BC}{ABC}\nonumber
\end{align}
These metrics are of Petrov type D and they have four Killing vectors and no homothety vector. Since $C_{\mu\nu\rho\sigma}C^{\mu\nu\rho\sigma}\propto \Sigma^{-12}$, these nonconformally-flat Einstein spaces are nonsingular as long as there are no points where $\Sigma=0$.

As another possible variation we could consider off-diagonal metrics, for instance with off-diagonal components having mixed products of first derivatives. There are solutions of this type, but we find that these solutions can be brought back into diagonal form via a coordinate transformation.

\section{Physics topics}
We have studied piecewise-Lorentzian manifolds involving three-dimensional hypersurfaces on which one component of the metric vanishes. The metric on and near these degeneracy planes is otherwise finite and smooth. More than one component of the metric vanishes at intersections of the degeneracy planes (in particular for solutions of Type 2, 3 and 4). The primary physical characteristic of these hypersurfaces is that all curvature invariants on and near them are finite. Then new classes of solutions to the Einstein equations can be constructed where the curvature invariants are everywhere finite. In this paper we have presented a variety of such solutions, either vacuum solutions or solutions sourced by a smooth matter distribution satisfying the standard energy conditions. Rather than summarizing these solutions here, let us turn to the physical implications of the degeneracy planes.

In \cite{Holdom:2023xip} we discussed the propagation of a particle through a spacelike degeneracy plane by considering the geodesics. For a particle travelling in the positive $x$ direction we obtained the quantities $dx/d\lambda$ and $dt/d\lambda$ by standard methods. From those results we get the apparent speed
\begin{align}
\frac{dx}{dt}\propto \sqrt{-g_\textit{tt}}\propto |t-t_0|
\end{align}
of a particle that is close in time to the plane situated at $t=t_0$. This results in
\begin{align}
x(t)-x_0\propto {\rm sign}(t-t_0)(t-t_0)^2
,\end{align}
and so the apparent speed is instantaneously zero when the particle crosses the plane at $x(t_0)=x_0$. For a timelike degeneracy plane we have
\begin{align}
\frac{dx}{dt}\propto \frac{1}{\sqrt{g_\textit{xx}}}\propto \frac{1}{|x-x_0|}
\end{align}
of a particle that is close to the plane situated at $x=x_0$. This results in
\begin{align}
x(t)-x_0\propto {\rm sign}(t-t_0)\sqrt{|t-t_0|}
,\end{align}
and so the apparent speed is instantaneously infinite when the particle crosses the plane at $x(t_0)=x_0$. In both cases $x(t)$ is continuous.

In \cite{Holdom:2023xip} we also found that solutions of the scalar wave equation were well behaved around a spacelike degeneracy plane such that the scalar $g^{\mu\nu}\partial_\mu\phi\partial_\nu\phi$ remained finite. This is also true for a timelike degeneracy plane.

How could piecewise-Lorentzian manifolds be involved in the description of the real world? Some of our solutions have the form of localized gravitational lumps, nonsingular solutions that are asymptotically flat. The vacuum solutions may be more interesting since the perfect fluid solutions require negative pressure. Although these solutions have spherically symmetric curvature invariants, they still display nontrivial structure asymptotically, as seen by the volume element in (\ref{e19}). This is a consequence of the degeneracy planes that extend out to infinity. How such an object could be singly produced by some local mechanism is not obvious.

The appearance of piecewise-Lorentzian manifolds in a cosmological context seems to be more promising. In \cite{Holdom:2023xip} we showed that a spacelike degeneracy plane represents a turning point for the cosmological scale factor, and thus it can model a nonsingular bounce cosmology with normal matter. Typically a violation of the null energy condition is required for a nonsingular bounce cosmology (for a review see \cite{Ijjas:2018qbo}). The only other proposal we are aware of that does not require such a violation is found in \cite{Klinkhamer:2019frj}, and this is also based on a metric defect that involves degeneracy. In other respects this construction differs from ours.

We have been describing the hypersurfaces as degeneracy planes, but the degeneracy property of the metric is coordinate dependent. By changing the coordinate system, the degeneracy of the metric can be traded for a nonsmooth but continuous behavior of the metric, as a hypersurface is crossed. We know how to construct the coordinate transformation that can accomplish this globally for solutions of Type 1, as described in \cite{Holdom:2023xip} for Type 1b. Thus the hypersurface is the location of a type of defect whose description is coordinate dependent. This is in contrast to the finiteness and the continuity of the curvature invariants, since these are physical properties that are independent of the coordinate system.

Let us return to the vacuum solutions of Type 1a or 1b that occur in the presence of a cosmological constant. These solutions can exhibit periodic oscillations that occur in space or time respectively. We may change coordinates such that the metric degeneracy is traded for nonsmooth behavior. Then the metric component that previously periodically vanished is transformed into a constant, either $-1$ or $+1$ as appropriate, and the function $d(t)$ or $a(x)$ is now a piecewise function switching between increasing and decreasing exponentials. The range of these functions matches the range of the original oscillations. There can be a reflection symmetry around each transition point. We thus have an infinite set of transitions in a layered structure that is perpendicular to either a space or time dimension. As we have mentioned, the nontrivial structure may be occurring on such short length or time scales so as to be effectively hidden from us.

A nonsmooth $Z_2$-symmetric transition would conventionally be described as an orbifold, as occurs for example along a fifth dimension in models of the Randall-Sundrum type. An orbifold normally requires specifying a brane with an appropriate tension situated at the point of nonsmooth behavior. In contrast for piecewise-Lorentzian manifolds no such externally added brane is needed; in our case the nonsmooth metric is obtained via a coordinate transformation of a smooth and oscillating metric that is sourced only by a cosmological constant. This indicates that another use of piecewise-Lorentzian manifolds may be in the formulation of higher-dimensional models.

%\acknowledgments

\end{document}